\documentclass[a4paper,11pt]{article}
\usepackage{jheppub} 
\usepackage{tikz}
\usepackage{slashed}

\title{\boldmath Diamonds in the Bulk and Large-$N$ Scaling in AdS/CFT}

\author[1]{Sidan A\note{Corresponding author.}}
\author{and Tom Banks}
\affiliation{Department of Physics and NHETC,
Rutgers University, \\
Piscataway, NJ 08854 USA}

\emailAdd{sidan.aa@rutgers.edu}

\abstract{ Quantum Field Theory (QFT) introduced us to the notion that a causal diamond in space-time corresponded to a subsystem of a quantum mechanical system defined on the global space-time. Work by Jacobson \cite{ted95}, Fischler and Susskind \cite{fs} and particularly Bousso \cite{bousso,bousso2,bousso3} suggested that in the quantum theory of gravity this subsystem should have a density matrix of finite entropy.  These authors formalized older intuitive arguments based on black hole physics.  Although mathematically, Type II von Neumann algebras admit finite entropy density matrices, the black hole arguments suggest that the number of physical states in these subsystems is finite.  The conjecture that de Sitter (dS) space has a finite number of physical states was first made in \cite{tb, wf}.  Leutheusser and Liu \cite{LL} showed that, in the $N = \infty$ limit, causal diamonds with finite area in AdS radius units had Type $III_1$ von Neumann sub-algebras of the full operator algebra. They claimed that this was true for finite values of the UV cutoff, and that the algebra was the algebra of bulk local fields in the diamond.  We will argue that the second part of this conjecture is incorrect and that the bulk field algebra emerges only in a double scaled limit, where the boundary UV cutoff is taken to infinity as $N$ is taken to infinity. There is never a bulk field theory description that resolves distances smaller than the AdS radius.  }

\begin{document}
\maketitle
\flushbottom

\section{Diamonds in Bulk}

In a remarkable series of papers \cite{LL} Leutheusser and Liu showed how to describe certain causal diamonds in the bulk of AdS space, using the language of algebraic quantum field theory (AQFT).  Their arguments take place in the strict $N = \infty$ limit, but are supposed to have a certain amount of validity to all orders in the expansion in inverse powers of $N$.\footnote{It is not exactly clear how this works. Hamilton, Kabat, Lifschytz, and Lowe (HKLL) \cite{HKLL,HKLL1} have systematized the all orders construction of bulk fields from boundary operators, which includes non-local gravitational Wilson lines. The tensor network construction was shown \cite{harlowetal,harlowetal1} to resolve paradoxes that arise if one assumes sub-algebras with non-trivial commutants in finite $N$ AdS/CFT. Are these paradoxes really avoided to all orders in the expansion just by including the leading order Hamiltonian in the algebra?}  When one of the present authors first read these papers, the following two issues immediately arose.
\begin{itemize}
\item Some of the arguments in \cite{LL} seem to be valid in {\it any} matrix quantum mechanics, yet the simple one matrix models do not have Type $III$ local subalgebras at large $N$, but only in the double scaled limit, where the boundary UV cutoff is taken to infinity as $N$ is taken to infinity. 
\item The Banks and Fischler \cite{tbwfads1} had identified finite causal diamonds (and implicitly, local subalgebras) in AdS/CFT for finite $N$ using the tensor network renormalization group \cite{tnrg,EV1}.  What was the relation between the two constructions?
\end{itemize}

The key to the construction in \cite{LL} was the isolation of Type $III_1$ {\it time band sub-algebras} of the $N = \infty$ algebra of single trace gauge invariant operators in the CFT.  These existed because, in a natural choice of energy scaling as $N \rightarrow \infty$, the CFT Hamiltonian $K_0 + P_0$ does not act on the algebra.  This is not the choice where this generator counts the dimensions of operators (and is thus dimensionless), but the bulk choice where it is multiplied by $R_{AdS}^{-1}$.  In addition, one is choosing to measure time in Planck units, so that a time of order $R_{AdS}$ goes to infinity in the limit.  At finite $N$ of course a famous theorem tells us that the fields in any arbitrarily small time slice generate the full Type $I_{\infty}$ algebra of the QFT.  

That algebra of course has many Type $III_1$ subalgebras, associated with finite causal diamonds on the boundary.  In a CFT with an Einstein-Hilbert dual, these are dual to infinite boundary anchored Ryu-Takayanagi \cite{RT} diamonds and have nothing to do with the finite bulk diamonds defined by \cite{LL}.  In order to clearly distinguish them, the authors of \cite{LL} invoke an unspecified UV cutoff, so that the finite $N$ cut-off CFT no longer has Type $III_1$ subalgebras\footnote{This is done on page 12 of their preprint \cite{LL} on subregion-subalgebra duality and is easily missed by the casual reader.}.  

The issue of UV cut-offs in the AdS/CFT correspondence is generally ignored by most of the community that works on the subject.  Enormous amounts of evidence, from many different kinds of calculations, show that UV divergences on the boundary are related to large volume divergences in AdS space, but the implications of this for the bulk to boundary correspondence have not really been appreciated.  As a consequence, we feel that it is necessary to add a small refresher section on renormalization to this paper.

\section{Ultraviolet Cutoffs and Conformal Field Theory}

K. Wilson taught us the true meaning of renormalization and continuum quantum field theory in the late 1960s.  There are an infinite number of different ways to cut-off a formal continuum classical field theory, even if we insist that the cut-off preserve unitarity.  Within a given cut-off scheme one can formulate renormalization group (RG) transformations, relating models with different cut-off scales.  CFTs are fixed manifolds of this RG semi-group flow and more general QFTs are constructed by taking double scaled limits along unstable directions from these fixed manifolds, as the length cut-off goes to zero.  Experience shows that one can generally map the models in one cut-off scheme to those in another, though there is no general theory of this mapping, especially far from the fixed manifolds. 

Thinking about this in the AdS/CFT context, one has to ask which cutoff scheme corresponds to the bulk physics of AdS space, or whether that physics is fundamentally ambiguous.  From this point of view, the constructions of \cite{HKLL,HKLL1} seem somewhat misleading, since they seem to give the impression that one can construct bulk physics from the continuum boundary.  We should note however that this is true only for the single trace operators, and that these do not account for most of the entropy in black holes, since the ensemble of single trace operators is dual to the AdS gas.  Moreover, Hamilton, Kabat, Lifschytz, and Lowe (HKLL) have shown that their expansion has ambiguities analogous to those encountered in extracting local field correlators from the S-matrix in ordinary QFT.  So the HKLL construction, which is central to the work of \cite{LL} is missing much of the quantum information about the structure of bulk AdS space.

One popular cut-off scheme is provided by the irrelevant $T\overline{T}$ perturbation \cite{TTbar,TTbar1,TTbar2,TTbar3,TTbar4}, which formally seems to pull the boundary in to a finite radius. In $AdS_3$, where this can be rigorously formulated as an integral perturbation of the CFT Hamiltonian, it leads to a Hamiltonian with complex eigenvalues.  The cut-off then consists in discarding eigenvalues above the point where they become complex.  This defines a quantum system on a continuous sphere of a fixed radius, which is effectively non-local in time.  The more one localizes the system in the radial direction, the greater the non-locality in time.  

The tensor network/Quantum Error Correcting Code (TN/QECC) cut-off of CFTs \cite{harlowetal,harlowetal1} of CFTs, is often described as {\it a toy model of the AdS/CFT correspondence}.  This is incorrect.  Tensor networks were introduced by condensed matter physicists as a systematic way to capture the entanglement structure of the ground state of many body systems, and CFTs in particular.  For simple models, the Tensor Network Renormalization Group (TNRG) \cite{tnrg,EV1} provides explicit machinery for calculating dimensions of operators at non-perturbative fixed points.  TN/QECC gives an explicit picture of a spatial slice of AdS in global coordinates and resolves paradoxes associated with naive bulk local field theory predictions.  It also gives the correct state counting for large AdS black holes, and a rigorous derivation of the Quantum Extremal Surface Conjecture \cite{harlowqec}.  It explicitly incorporates Maldacena's scale/radius duality.  In \cite{tbwfads1} we showed that the embedding maps of the TNRG could be viewed as analogs of modular inclusion in large causal diamonds in AdS space, with successive shells of the tensor network playing the role of the maximal volume surface on the boundaries of a nested sequence of causal diamonds along the central geodesic in the global coordinates.  Thus, the TN/QECC cut-off is the most plausible answer to the question of which cut-off of a holographic CFT mirrors the correct bulk physics.  

In a holographic CFT (which we call a CFT with an Einstein-Hilbert dual), each node of the tensor network contains many q-bits and represents a large region of space.  We believe that this region should be thought of as one whose domain of dependence is the causal diamond that Polchinski and Susskind called {\it the arena} in their seminal papers on the derivation of Minkowski scattering amplitudes from CFT correlators.  This is a region whose size scales to infinity with $N$, but is parametrically smaller than the AdS radius by a small power of $N$.

\section{Diamonds in the Bulk in a Large N Tensor Network}

Given this preamble, we choose to use the TN/QECC cut-off scheme.  We do this in the following way in $AdS_d$:

\begin{itemize}
\item On the hyperbolic space $H^{d-1}$ place a ball of radius $R$. Eventually we'll choose $R$ slightly smaller than $R_{AdS}$, but scaling to infinity as $N$ goes to infinity, so that it is larger than any microscopic scales. We can imagine something like 
\begin{equation}
    R = R_{AdS} \left(\frac{R_{AdS}}{L_P}\right)^{-\epsilon},
\end{equation}
where the small number $\epsilon$ will depend on the particular model. If we think of the domain of dependence of the ball $R$ as the Polchinski-Susskind ``arena" \cite{polchsuss,polchsuss1}, then the size of $\epsilon$ depends on how much Minkowski space physics we want to be captured accurately by CFT calculations.
Now fill the rest of the space with close packed balls of radius $R$, centered around the original one.  The centers of all those balls form a lattice on the hyperbolic space, invariant under a discrete subgroup of its isometries.  The shells at fixed radial distance from the center in lattice steps form lattices on $S^{d-2}$. See Figure \ref{fig:1}.

\begin{figure}[htbp]
\centering
\includegraphics[width=.41\textwidth]{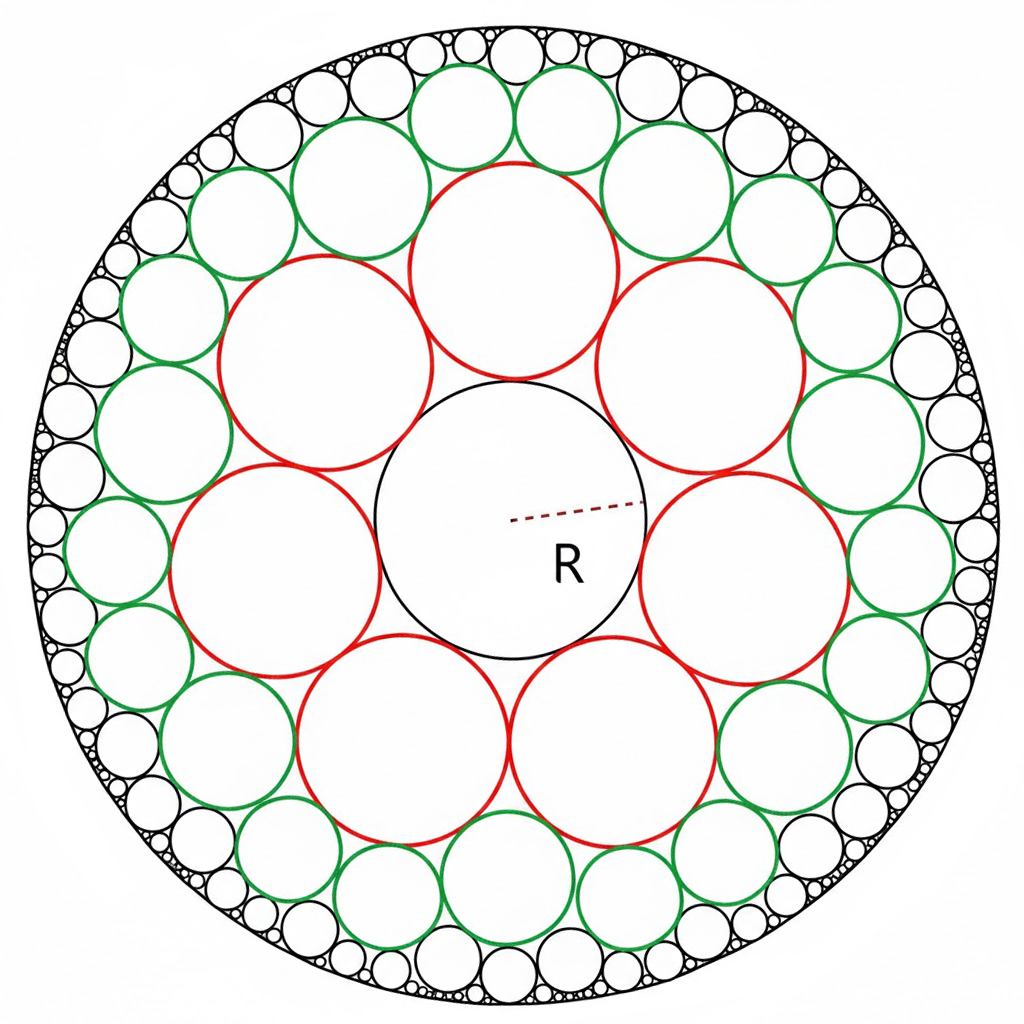}
\qquad
\includegraphics[width=.4\textwidth]{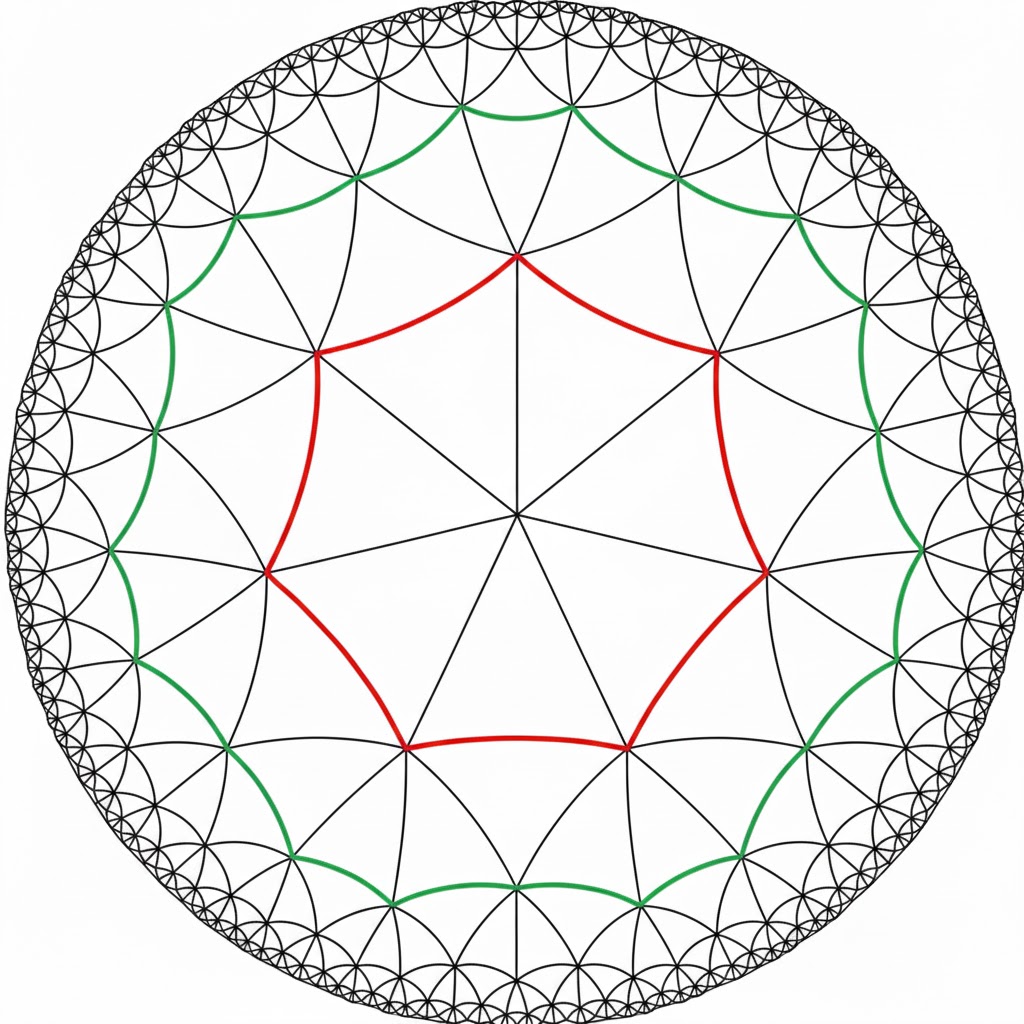}
\caption{An $d=3$ example of the packing on the hyperbolic space $H^2$. Each ball in the diagram on the left has the same radius $R$, and their centers form a lattice on the hyperbolic space shown on the right. $S^1$ at one and two lattice steps from the center are also shown in the diagram on the right in red and green, respectively. The red shells on the left form a lattice on the red $S^1$ on the right, and the green shells on the next outer layers form a lattice on the green $S^1$ one lattice step further from the center. 
\label{fig:1}}
\end{figure}

\item We are going to try to build a Tensor Network Renormalization Group (TNRG) \cite{tnrg,EV1} which maps each lattice point on a given shell into those points nearest it on the next larger shell. The Hilbert space on each lattice point will have the same dimension, which will go to infinity like the exponential of a power of $R$.  We view the points at the centers of a given shell to be lying on the bifurcation surface of a causal diamond of fixed proper time along the AdS geodesic that sits at the central point, with time reflection symmetry about our hyperbolic space. That is to say, the center of the green $S^1$ in Figure \ref{fig:1} is on a bifurcation surface at later time than the bifurcation surface on which the center of the red $S^1$ lies. Match the geometric entropy of these surfaces to the number of lattice points on the surface times the logarithm of some integer.  We adjust the geometric size of the ball at the central point to make all these calculations come as close as possible to integers.
\item We assume that we know the exact spectrum of the CFT and match the size of the single site Hilbert space to the size of the Hilbert space obtained by acting on the CFT vacuum with all operators (primary and descendant) below some dimension $\Delta_*$. The single site Hamiltonian is chosen to coincide with the operator dimension counting CFT Hamiltonian $K_0 + P_0$, up to a scale factor we'll discuss below.  However, since we are discussing the site at the center of the lattice, we omit descendants that move the operator away from that site.  So the descendants are constructed using only the subgroup of the conformal group that commutes with the isometries of the hyperbolic slice.  This subgroup is $SO(2)$, generated by the Hamiltonian.  
\item  Following \cite{happy} we imagine that we can find a sequence of local embedding maps of one shell of the tensor network into the next, that is invariant under a discrete subgroup of the rest of the AdS isometry group. This subgroup maps the centers of different close packed spheres into each other.  Based on the results of Evenbly and Vidal \cite{tnrg, EV1} in exactly soluble models, we conjecture that the complete lattice Hamiltonian generated on each shell of the network converges to the $K_0 + P_0$ generator of the CFT.  The convergence should be such that lower eigenvalues of the field theory Hamiltonian are captured more accurately on smaller shells, implementing Maldacena's scale radius duality.   We have started out with a Hamiltonian for the central node whose eigenvalues are the low lying dimensions.  
\item The resulting sequence of lattice theories realizes the idea of Swingle \cite{swingle, swingle1} that tensor networks capture the bulk to boundary map of the AdS/CFT correspondence.
\item In \cite{tbwfads1} we argued that the shells of the tensor network should be interpreted as the boundaries of causal diamonds along a geodesic running through the center of the global coordinate system defined by $K_0 + P_0$ and that this converted the embedding maps of the TNRG into a discrete analog of causal time evolution (a sort of two side modular inclusion) in the proper time of that geodesic. 

\end{itemize}

It is clear from these constructions that if we take a very coarse grained lattice, the argument of \cite{LL} that the $N = \infty$ limit of the algebra of single trace operators in the CFT is the algebra of fields in a bulk CFT in AdS space is not correct.  For the central node of a tensor network construction of the CFT it is the algebra of that node, and for finite shells of the network it is a finite tensor product of central node algebras. That algebra contains no excitations of non-zero orbital angular momentum on $S^{d-2}$.  In order to gain some understanding of the central node algebra let us restore the finite energies/dimensions of the operators we have included in the node Hilbert space.  The energies of the low lying operators that have been included in the single node Hilbert space are all of order $R_{AdS}^{-1}$ if we view the CFT as compactified on a sphere of radius equal to the AdS radius.  

For CFTs that have an Einstein-Hilbert dual, the single node corresponds to a spatial region on a slice of fixed global time in the coordinate system defined by $K_0 + P_0$, whose domain of dependence is roughly the causal diamond that Susskind called ``the arena" in his derivation of Minkowski S-matrix elements from AdS correlators \cite{polchsuss, polchsuss1}.  We see an immediate contradiction with the idea that time evolution generated on the single node algebra by $K_0 + P_0$ is related to bulk time evolution in that diamond, as defined by the bulk Hamiltonian in any gauge one might choose for gravitational effective field theory in $AdS_D \times {\cal K} $.  ${\cal K}$ is a $k$ dimensional compact manifold, whose linear length scales are of order the AdS radius, $R_{AdS}$.  

It's clear that the lattice cutoff does not preserve rotations on $S^{d - 2}$, but we can organize the operators in the single node Hilbert space into $SO(d - 1)$ multiplets, because that sphere {\it is} invariant.  If we succeed in finding tensor network maps that preserve a discrete subgroup of $SO(d - 1,2)$ that maps the centers of closely packed spheres into each other, then this classification of multiplets can be transported to all the nodes of the network.  However, it is clear that only zero orbital angular momentum components of local operators should appear in the single node algebra.  Spatial variations over $S^{d - 2}$ correspond to jumping between lattice points and the full continuous $SO(d - 1)$ rotation symmetry is only recovered on the boundary.  

As $N \rightarrow \infty$ in a single node, the maximal subgroup of $SO(d - 1, 2)$ that can be recovered is the $SO(2)$ acting as translation on the $t$ coordinate of a global coordinate system.    Indeed, the zero angular momentum equations for massive particles are identical to their asymptotic forms for any mass, so $SO(2)$ is a symmetry of the radially symmetric part of the full effective field theory.  

\subsection{Scalar fields}
Let us examine the radial equations in the near horizon region of a causal diamond whose size is small compared to the AdS radius. The equation of motion for a free scalar field in AdS space is given by
\begin{equation}\label{eq:scalarAdS}
    \begin{split}
        \left\{-\frac{\partial_{\tau}^2}{1+r^2/R_{AdS}^2}+\left(1+\frac{r^2}{R_{AdS}^2}\right)\partial_r^2+\frac{1}{r}\left[d\left(1+\frac{r^2}{R_{AdS}^2}\right)-2\right]\partial_r-m^2\right\}\phi &= 0 .
    \end{split}
\end{equation} 
We have taken the $N = \infty$ limit, so the bulk fields satisfy free field equations.  We can write 
\begin{equation}
    \phi = f(r) \chi (\tau,r) .
\end{equation} 
The $\partial_r\chi$ term vanishes when
\begin{equation} 
    2\left(1+\frac{r^2}{R_{AdS}^2}\right)\partial_r f +\left[\frac{d}{r}\left(1+\frac{r^2}{R_{AdS}^2}\right)-\frac{2}{r}\right]f = 0,
\end{equation}  
then $\chi(\tau,r)$ satisfies 
\begin{equation} 
    \begin{split}
        -\frac{\partial_{\tau}^2 \chi}{1+r^2/R_{AdS}^2} + \left(1 + \frac{r^2}{R_{AdS}^2}\right) \partial_r^2 \chi + F(r)\chi = 0 ,
    \end{split}
 \end{equation}  
where
\begin{equation}
    F(r) \equiv \left(1 + \frac{r^2}{R_{AdS}^2}\right) \frac{\partial_r^2 f}{f} + \left[\frac{d}{r}\left(1 + \frac{r^2}{R_{AdS}^2}\right) - \frac{2}{r}\right]\frac{\partial_r f}{f} -m^2 .
\end{equation}
For $ r $ sufficiently small compared to $R_{AdS}$ and near the boundary of the causal diamond of size $R$, one of the light front coordinates $x^{\pm} = \frac{1}{2}(r\pm t)$  is very small, say $x_+ \ll x_-$. Then we have $\partial_+\gg \partial_-$. This implication is not true in general, but we are thinking of fields confined to a small region near the light cone, whereas the other light-like parameter varies freely over the entire cone, so if the field is confined to a region where $x_+\ll x_-$, it contains Fourier modes of both long and short wavelengths in the $x_-$ direction, but in the $x_+$ direction the field has only Fourier modes of short wavelengths, therefore $\partial_+ \gg \partial_-$. Then rewrite the above equation in light front coordinates, we find that $\chi(\tau,r)$ satisfies
\begin{equation} 
    2 \partial_+ \partial_- \chi + \mathcal{O}(1) \chi = 0 . 
\end{equation}
The order one terms are negligible and the near horizon quantum variable is a massless free field. These terms are order one because they depend only on the variable $r$. Thus, taking the central node of the tensor network to be large but smaller than the AdS radius, e.g. 
\begin{equation}
    R = R_{AdS} \left(\frac{R_{AdS}}{L_P}\right)^{-\epsilon},
\end{equation}
for some small positive $\epsilon$, one finds that one should match its bulk entropy by quantizing the bulk field dual to each spin zero primary as a one plus one dimensional CFT on the stretched horizon of the central diamond, with $c = 1$. As similar result is found for the transverse components of vector and tensor fields.

\subsection{Dirac fields}
Let us first consider a free Dirac spinor $\psi$ in AdS space, which has the following equation of motion
\begin{equation}
    \begin{split}
        \left[i\gamma^ke_k^{\mu}\left(\partial_{\mu}+\frac{1}{4}\omega_{\mu ij}\gamma^{ij}\right)-m\right]\psi &= 0,
    \end{split}
\end{equation}
where spin connections $\omega_{\mu ij}$ can be found using the vielbein $e_{\mu}^i$, and the second rank Clifford matrices is defined as
\begin{equation}
    \begin{split}
        \gamma^{ij} &= \frac{1}{2}[\gamma^i, \gamma^j] .
    \end{split}
\end{equation}
In AdS space, the equation of motion is given by
\begin{equation}
    \begin{split}
        \left[\frac{i\gamma^0 \partial_{\tau}}{\sqrt{1+r^2/R_{AdS}^2}}+i\gamma^1\sqrt{1+\frac{r^2}{R_{AdS}^2}}\left(\partial_r+\frac{d-1}{2r}-\frac{1}{2r\left(1+r^2/R_{AdS}^2\right)}\right)+ \frac{i}{r}\slashed{D}_{S^{d-2}}-m\right]\psi &= 0,
    \end{split}
\end{equation}
where the Dirac operator on the unit $m$-sphere can be expressed as
\begin{equation}
    \begin{split}
        \slashed D_{S^{m}} &= \gamma^{\theta_1}\left(\partial_{\theta_1}+\frac{m-1}{2}\cot\theta_1\right) + \frac{1}{\sin\theta_1}\slashed D_{S^{m-1}} \\
        &=\sum_{k=1}^m\frac{\gamma^{\theta_k}}{\prod_{n=1}^{m-1}\sin\theta_n}\left(\partial_{\theta_k}+\frac{m-k}{2}\cot\theta_k\right) .
    \end{split}
\end{equation}
When $r\ll R_{AdS}$, the equation of motion reduces to
\begin{equation}
    \begin{split}
        \left[i\gamma^0\partial_{\tau} + i\gamma^1\left(\partial_r + \frac{d-2}{2r}\right) + \frac{i}{r}\slashed{D}_{S^{d-2}} -m\right]\psi &=0 .
    \end{split}
\end{equation}
We can drop the angular term since only zero orbital angular momentum components should appear due to the lack of rotational symmetries, we have
\begin{equation}\label{eq:diracflat}
    \begin{split}
        \left[\gamma^0\partial_{\tau} + \gamma^1\left(\partial_r + \frac{d-2}{2r}\right)-m\right]\psi &= 0 . \\
    \end{split}
\end{equation}
Separate the Dirac field as $\psi(\tau,r) = f(r)\chi(\tau,r)$,
\begin{equation}
    \begin{split}
        \text{if} \hspace{2mm} \left(\partial_r +\frac{d-2}{2r}\right)f =0, \hspace{10mm} \text{then} \hspace{2mm} \left(\gamma^0\partial_{\tau}+\gamma^1\partial_r-m\right)\chi &=0.
    \end{split}
\end{equation}
Again the mass term is order one and can be negligible, and we get the 1+1 dimensional Dirac equation.

\subsection{Vector fields}
For a free spin-1 field, its equation of motion is given by the Maxwell's equation in curved spacetime 
\begin{equation}
    \begin{split}
        \nabla_{\mu}F^{\mu\nu} - m^2A^{\nu}=0 , \\
    \end{split}
\end{equation}
where $F_{\mu\nu} = \nabla_{\mu}A_{\nu} - \nabla_{\nu}A_{\mu}$. Note that we use $\nabla$ notation for covariant derivative here to emphasize that we only have affine connections $\Gamma^{\mu}_{\nu\rho}$, with it acts on spin-1 vector fields as $\nabla_{\mu}A_{\nu} = \partial_{\mu}A_{\nu} - \Gamma^{\rho}_{\mu\nu}A_{\rho}$. Using 
\begin{equation}
    [\nabla_{\mu}, \nabla_{\nu}]A^{\rho} = R^{\rho}{}_{\sigma\mu\nu}A^{\sigma},
\end{equation}
and we find 
\begin{equation}\label{eq:spin1eom}
    \begin{split}
        \nabla^2A^{\nu} - g^{\alpha\nu}R_{\rho\alpha}A^{\rho} - m^2A^{\nu} & = 0.
    \end{split}
\end{equation}
In AdS space, we have the following equation of motion for each component
\begin{equation}\label{eq:spin1ads}
    \begin{split}
        \forall \hspace{1mm} \nu \in \{\tau,r,\theta_1, \cdots, \theta_{d-3}\} \hspace{5mm} \Rightarrow \hspace{5mm}\left(\nabla^2+ \frac{d-1}{R_{AdS}^2} - m^2 \right)A^{\nu} &=0 .
    \end{split}
\end{equation}
When $r\ll R_{AdS}$, the equation of motion reduces to the case for flat spacetime. In flat spacetime, the Ricci tensor is just a zero tensor, so the second term in Eq.(\ref{eq:spin1eom}) vanishes. Following the same argument, the angular term can be ignored and we get
\begin{equation}\label{eq:spin1flat}
    \begin{split}
        \left(-\partial_{\tau}^2  + \partial_r^2 + \frac{d-2}{r}\partial_r - m^2\right)A^\nu &= 0.
    \end{split}
\end{equation}
This is the same equation as for scalar fields, separate the fields the same way as before with $A^{\mu}(\tau,r) = f(r)\chi^{\mu}(\tau,r)$, we have
\begin{equation}
    \begin{split}
        \text{if} \hspace{2mm} \left(\partial_-+\frac{d-2}{2x^-}\right)f = 0, \hspace{10mm} \text{then} \hspace{2mm} \partial_+\partial_-\chi^{\nu} & =0 .
    \end{split}
\end{equation}

\subsection{Tensor fields}
The equation of motion for a free massive spin-2 field is given by
\begin{equation}\label{eq:spin2eom}
    \begin{split}
        \nabla^2h^{\mu\nu}-\partial^{\nu}\partial_{\rho}h^{\mu\rho} -\partial^{\mu}\partial_{\rho}h^{\nu\rho}+ \partial^{\mu}\partial^{\nu}h + \eta^{\mu\nu}\partial_{\rho}\partial_{\sigma}h^{\rho\sigma} -\eta^{\mu\nu}\nabla^2h-m^2\left(h^{\mu\nu}-\eta^{\mu\nu}h\right) &= 0.
    \end{split}
\end{equation}
Taking $\partial_{\mu}$ on both sides of the above equation gives
\begin{equation}
	\begin{split}
		m^2\left(\partial_{\mu}h^{\mu\nu}-\partial^{\nu}h\right) &= 0, \\
	\end{split}
\end{equation}
and for massive tensor fields, this reduces to
\begin{equation}\label{eq:spin2relation}
    \begin{split}
         \partial_{\mu}h^{\mu\nu} - \partial^{\nu}h &=0.
    \end{split}
\end{equation}
Now take the trace of both sides of Eq.(\ref{eq:spin2eom}), and we get
\begin{equation}
	\begin{split}
		m^2(d-1)h &= 0
	\end{split}
\end{equation}
Since $m\neq 0$ for massive fields, the fields are traceless for any $d>1$ and bringing it back to Eq.(\ref{eq:spin2relation}), we find the fields to be transverse, i.e.
\begin{equation}
    \begin{split}
        h=0, \hspace{10mm} \text{and} \hspace{10mm} \partial_{\mu}h^{\mu\nu}=0.
    \end{split}
\end{equation}
Bringing the transverse-traceless condition back to the equation of motion, we find
\begin{equation}
    \begin{split}
        \left(\nabla^2-m^2\right)h^{\mu\nu} &=0,
    \end{split}
\end{equation}
which is the same as the scalar case, thus the same reasoning applies and letting $h^{\mu\nu}(\tau,r) = f(r)\chi^{\mu\nu}(\tau,r)$, we conclude that
\begin{equation}
	\begin{split}
		\text{if} \hspace{2mm} \left(\partial_-+\frac{d-2}{2x^-}\right)f = 0 \hspace{3mm} \text{then} \hspace{2mm} \partial_+\partial_-\chi^{\mu\nu} & =0 .
\end{split}
\end{equation}

We've thus effectively derived the conjecture of \cite{carlip, solo, BZ} in the $N = \infty $ limit of AdS/CFT models, by assuming the tensor network regularization of the CFT.  The conjecture was derived for diamonds in AdS space that are of the order of, but somewhat smaller than, the AdS radius.  A global spatial slice of maximal volume through such a diamond is the central node of the network.  

This construction is consistent with the results of \cite{VZ2, deBoeretal,deBoeretal1} that the fluctuation relation  
\begin{equation} 
    \langle (K - \langle K \rangle)^2 \rangle  = \langle K \rangle , 
\end{equation} 
is valid for the modular Hamiltonian of diamonds in CFTs that have Einstein-Hilbert duals.  In the bulk dual picture, $K$ is the modular Hamiltonian of the Ryu-Takayanagi (RT) diamond anchored to the boundary diamond, and the boundary UV divergence of both sides of this equation translates into the area divergence of the RT diamond measured in AdS radius units.  The coefficients of the divergent term represent the fluctuation in a diamond with size of order the AdS radius.  The results of \cite{VZ2, deBoeretal,deBoeretal1} show that that diamond obeys the Carlip-Solodukhin fluctuation law \cite{carlip,solo}.   

These results are also consistent with old arguments that bulk QFT cannot be a good approximation to quantum gravity in causal diamonds smaller than the AdS radius.  Most of the states implied by QFT in a causal diamond of size $R$, would back-react in semi-classical gravity to form a black hole larger than $R$.  Leaving out those states\footnote{The authors of \cite{CKNetal,CKNetal1,CKNetal2} showed that leaving out those states does not interfere with the precise agreement between QFT and experiment.}  leaves over an entropy that scales like 
\begin{equation}
    \left(\frac{R_{AdS}}{L_P}\right)^{\frac{(d - 1)(d-2)}{d}}, 
\end{equation}
not enough to account for the Bekenstein-Hawking-Jacobson-Fischler-Susskind-Bousso \cite{ted95, fs, bousso, bousso2, bousso3, bek,hawk} (BHJFSB) entropy of the diamond.  The attempt to account for black hole entropy in terms of QFT entanglement entropy on the horizon fails for multiple reasons \cite{thooftisrael,thooftisrael1,thooftisrael2,thooftisrael3,thooftisrael4, firewall}.  

Three aspects of the proposal in \cite{hilbertbundles,hilbertbundles1} have not been derived from AdS/CFT models.  The first is the claim that the near horizon CFT is constructed from free fermions in one to one correspondence with the (cut-off) spectrum of eigenspinors of the Dirac operator on the holographic screen of the diamond.  The second is that the CFT should actually be an interacting theory, in order to capture the fast scrambling property of horizons. The third, and the most important, is that the CFT obtained through tensor network approximation with the large N formalism of AdS/CFT fails to be a CFT with a gravitational dual.

The absence of the second property is clearly related to our study of the strict $N = \infty$ limit.  The boundary CFT is an integrable system in that limit, and cannot exhibit fast scrambling since all connected $k > 2$ point functions vanish.   Fast scrambling should appear once $1/N$ corrections are taken into account properly.   At any rate, it now appears that, within the context of the Carlip-Solodukhin ansatz, fast scrambling is implemented by marginally irrelevant perturbations of the free fermion CFT, rather than an interacting CFT \cite{toappear}.  

At the level of free fields, we can of course use bosonization techniques to write everything in terms of fermions, but a more principled derivation would use the fact that low lying operators of the boundary CFT all lie in BPS multiplets, so that it should be possible to obtain them by acting with fermionic operators only.  We will reserve the study of this question for future work. 

One might object to the arguments above because they relied on a lattice cutoff of the CFT.  Perhaps the picture of \cite{LL} is valid at finite cutoff with some other cutoff scheme.  Recall however that CFT, and QFT more generally is defined to be independent of the cutoff scheme.  Thus, if we believe in Maldacena's scale/radius duality as expressed in \cite{susswit}, finite causal diamonds are not unambiguously represented in the boundary CFT.  A tensor network/error correcting code construction of the bulk, as proposed in \cite{harlowetal,harlowetal1, tbwfads1, happy, swingle, swingle1} provides an explicit sequence of cutoff models, which guarantee a coarse grained version of unitary causal bulk time evolution.  If our conjectured generalization of the construction of \cite{happy} is valid, then the full network is invariant under a discrete subgroup of the isometry group of AdS space and is designed to converge to the CFT in the infinite volume limit.  It incorporates the BHJFSB area law for causal diamonds, and, is local only in the AdS dimensions of space, and on scales larger than the AdS radius.  As we will see in the next section, this means that the results of \cite{LL} only arise in a double scaling limit in which we take the size of the tensor network shell to scale to infinity like a power of $N$.  An alternate cutoff scheme, which avoided such a double scaling limit, would, of necessity have to probe locality in the CFT on distance scales short compared to the AdS radius, and would claim that a QFT description of bulk causal diamonds was valid on those scales.  There are reasons \cite{CKNetal,CKNetal1,CKNetal2} to believe that this is not the case. 

However, the most serious issue with our ``derivation" of the Carlip-Solodukhin prescription from the standard large N formalism of AdS/CFT combined with tensor networks, is that we are not able to account for the large value of the central charge $c$\footnote{This was pointed out to one of us (T.B.) by D. Harlow when the material in this paper was presented at an informal seminar at MIT.}.   The reason for this is that for holographic CFTs, models with an Einstein-Hilbert dual, there are low energy states of the emergent Minkowski scattering theory in the central node of the network, which do not correspond to limits of low energy states in the CFT.  This was pointed out in \cite{tbwfads2}, where it was argued that the indefinitely large number of soft gravitons in flat space scattering amplitudes arose from multipoint connected correlators in the CFT, in which {\it most} of the boundary operators create bulk excitations that do not enter the Polchinski-Susskind ``arena" diamond, whose boundary converges to conformal infinity in Minkowski space.  The surviving states, which appear to a detector on the central geodesic as low energy states living on the boundary of the arena, are, from the boundary CFT point of view, complicated linear combinations of energy eigenstates, including high energy states.  Recall that high energy scattering can produce large number of low energy gravitons.  

Thus, the simple lesson extracted from the work of Evenbly and Vidal from soluble CFTs, that the spectrum of small lattices in a tensor network approximation to a CFT is a good approximation to the low lying spectrum of the CFT, does not really extend to CFTs with Einstein-Hilbert duals.  The bulk of low energy states on small lattices are soft graviton states, which arise from complicated entanglement patterns with multipoint correlators outside the central node.  These can include entanglement with states created by high dimension operators that remain outside the arena but exchanges soft gravitons with it. Furthermore the entropy of the central node comes from terms that are essentially non-perturbative in the $1/N$ expansion, so it does not make sense to think of it as being captured by the $1/N$ corrections.  

\section{A Double Scaling Limit - Beyond the Lattice}

We have emphasized the TNRG sequence of lattice cutoffs of the boundary CFT because we think that it preserves as much as possible of the local unitary structure we expect to emerge from quantum gravity.  It is striking that it seems to lead to the generalized Carlip-Solodukhin prescription for the description of the density matrix of diamonds below the AdS radius scale.  However, the necessity of scaling the UV cutoff to infinity as $N \rightarrow\infty$ follows from much more general considerations.

The authors of \cite{susswit} argued long ago that bringing the AdS boundary in from infinity was equivalent to a UV cutoff on the CFT.  The entire literature on holographic renormalization, $T\overline{T}$ deformations and recent work \cite{evaetal,evaetal1,evaetal2,evaetal3,evaetal4,evaetal5,evaetal6} on boundary conditions on finite time-like surfaces in general relativity is premised on this connection.
Near the boundary of AdS, the metric is
 \begin{equation} 
    ds^2 = - \frac{e^{2\alpha}}{4} dt^2 + R_{AdS}^2 \left(d\alpha^2 +  \frac{e^{2\alpha}}{4}d\Omega_{d-2}^2\right) . 
\end{equation} 
The boundary CFT coordinates are $t$ and $\Omega$, and the natural energy scale is $R_{AdS}^{-1}$.  Leutheusser and Liu consider the large $N$ limit where this goes to zero like a power of $N$ and argue that in that limit, the time band sub-algebra for bands of finite $\Delta t$ has non-trivial commutant in the algebra of all operators whose dimension stays finite in the limit.  This commutant is identified with the algebra of operators in the diamond whose bifurcation surface coincides with the surface where light rays from the two ends of the time interval meet.  

\begin{figure}[htbp]
\centering
\includegraphics[width=.41\textwidth]{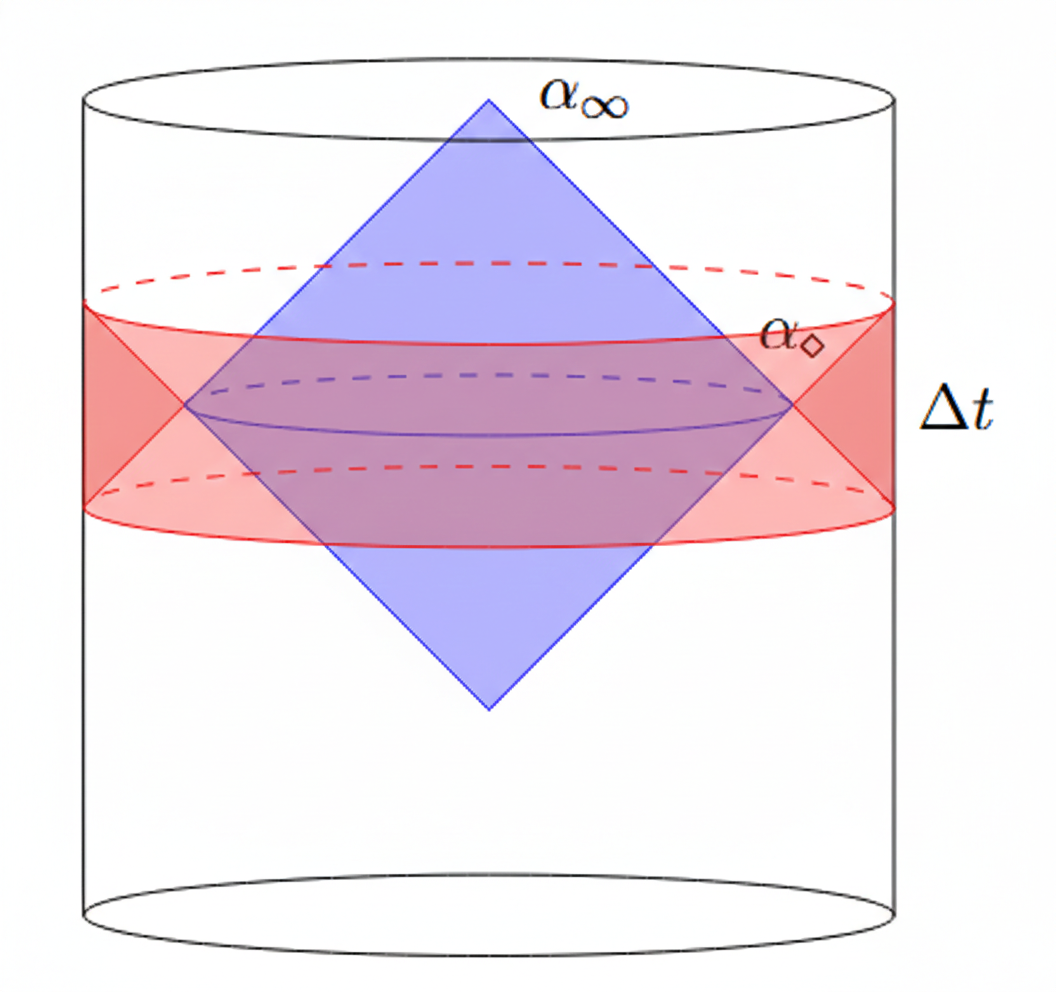}
\caption{A causal diamond (blue) in the global AdS. $\alpha_{\diamond}$ is where the wedge region (red) touches the casual diamond. $\alpha_{\infty}$ is the future tip of the causal diamond when we impose a UV cutoff. } 
\label{fig:2}
\end{figure}

If we impose a UV cutoff by imposing Dirichlet boundary conditions on the fields at some finite $\alpha_{\infty}$ deep in the asymptotic region, then an easy calculation shows that the bifurcation surface of the diamond lies at
\begin{equation} 
    e^{-\alpha_{\diamond}} = e^{- \alpha_{\infty}} + \frac{\Delta t}{4 R_{AdS}} . 
\end{equation}  
where $\alpha_{\diamond}$ and $\alpha_{\infty}$, and the time band $\Delta t$ are shown in Figure \ref{fig:2}. Thus, if we want the diamond to remain in the bulk for all finite times we must have
\begin{equation} 
    e^{\alpha_{\infty}} = \frac{r_{\infty}}{R_{AdS}} > \frac{R_{AdS}}{t} , 
\end{equation} 
for all finite $t$.  Thus, if we accept scale/radius duality, the Leutheusser and Liu proposal only makes sense in a double scaled limit in which we take the cutoff to infinity with $N$, independently of how the cutoff is defined.  

\subsection{The Polchinski-Susskind Arena}

The papers of \cite{polchsuss,polchsuss1} first established the beginnings of a strategy for extracting a non-perturbative formulation of quantum gravity in asymptotically flat space using the reliable techniques of AdS/CFT.  Using the language of Susskind, one focuses on a causal diamond of size $ L \ll R_{Arena} \ll R_{AdS} $ where $L$ is any microscopic length scale like $L_P$ or $L_S$, and prepares operators on the AdS boundary, which, in the Witten diagram approximation, produce a finite number of excitations that propagate freely until they enter the Arena diamond.  The resulting correlator produces, in the $R_{AdS}/L \rightarrow\infty$ limit, a Minkowski space S matrix element with a finite number of incoming and outgoing particles.  The number is restricted by the requirement that one does to put enough energy into the arena to create a black hole as large as the arena and by the fact that even massless particles carry energy of order $1/R_{AdS}$ in AdS space.  Amplitudes with arbitrarily large numbers of soft massless particles in Minkowski space come from connected CFT correlators containing a finite number of Polchinski-Susskind operators and an arbitrarily large number of operators that create excitations outside the arena but exchange gravitons with those in the arena in Witten diagrams.  In addition, the program never dealt with scattering amplitudes involving momentum components in the new non-compact directions that open up in the limit of all known AdS/CFT models. 

Our emphasis here is on the implications of the Polchinski-Susskind arena picture for the Leutheusser and Liu claims about operator algebras.  If we believed that the Leutheusser and Liu argument worked at the scale of the PS arena, then we would be claiming that the $R_{AdS}/L_P \rightarrow\infty$ limit of the area diamond algebra was the algebra of free quantum fields on the Penrose diamond of Minkowski space, contradicting the claims of Polchinski and Susskind.  The simple computation we did in the first part of this section and the more speculative tensor network based argument of the bulk of this paper shows why this cannot be correct. $1/N$ corrections to the HKLL construction will restore the agreement with perturbative calculations of the scattering matrix in bulk field theory (although the precise way in which the bulk UV divergences are regulated has not been completely clarified).  However, we have emphasized that this does not mean that there is any simple connection between the $1/N$ corrected algebra of HKLL fields and the entropy of finite causal diamonds in the bulk. Witten diagrams show that there is soft graviton entanglement between the Polchinski-Susskind arena and states created by operators of arbitrarily high dimension far outside the arena, which is not captured by the single trace operator algebra.  Simply adjoining the low energy Hamiltonian to the the algebra by the crossed product construction does not take these effects into account.  Only the TN/ECC formalism interprets the relationship between bulk field theory and the boundary correctly.

\section{Conclusions}

The key issue addressed by the considerations in this note is the nature of locality on scales smaller than the AdS radius in the AdS/CFT correspondence.  All of the work using error correcting codes, tensor networks, Ryu-Takayanagi diamonds, {\it etc.} only establishes locality on scales larger than or equal to the AdS radius.  This is demonstrated for example by its inability to make local probes of the compact extra dimensions with AdS scale radii, that accompany all known models.  Ironically it's also demonstrated by the fact that all of those ``local" features of CFTs are shared by CFTs that don't have large radius duals.  On the other hand, all of the follow up work to  \cite{LL} appears to assume QFT is a good approximation to quantum gravity on much smaller scales, contradicting well known arguments dating back to at least the 1980s.  Those arguments suggest that in flat space-time, quantum field theory fails to account for most of the entropy in a causal diamond.  The important paper of \cite{CKNetal,CKNetal1,CKNetal2} showed that this was entirely compatible with the success of QFT in reproducing all extant experimental data.  Conjectures about how QFT emerges from a correct theory of QG on scales smaller than the AdS radius can be found in \cite{hst,hst1,hst2}.

The point of the present paper was to demonstrate, at least within the framework of lattice approximations to CFT, that the crucial issue for resolving this point is the relation of the boundary UV cutoff and $N$, as $N$ is taken to infinity.  A boundary UV cutoff is equivalent to a volume cutoff on global spatial slices of AdS.  If field theory is a good approximation on scales small compared to the AdS radius, then the arguments of  \cite{LL} should be valid in the presence of a more or less arbitrary cutoff, while if it's only valid above that radius then the reason must be because there are many regions of AdS radius size near the boundary.  Only a correlated and double scaled limit can make the arguments of \cite{LL} into a systematic expansion of the theory.  

It should be pointed out that if the model of \cite{carlip, solo, BZ} for the dynamics on small diamond boundaries are correct then $1 + 1$-dimensional field theory {\it is} the correct description of bulk dynamics in $1 + 1$-dimensional models of QG.  It has a cutoff related to the total entropy in a diamond.  This is outlined in \cite{1dmodels,1dmodels1,1dmodels2}.  The relation of the ideas of \cite{LL} to large-$N$ limits in matrix models of $ 1 + 1$-dimensional QG was explored in \cite{tbdoublescaled}. 


\acknowledgments

The work of S. A is supported in part by the DOE under grant DE-SC0010008. T.B. thanks D. Harlow for an illuminating comment on this material during a talk at MIT.





\end{document}